# Performance of ChatGPT on the US Fundamentals of Engineering Exam: Comprehensive Assessment of Proficiency and Potential Implications for Professional Environmental Engineering Practice

Vinay Pursnani[a], Yusuf Sermet [a]*, and Ibrahim Demir [a,b,c]

[a] IIHR – Hydroscience & Engineering, University of Iowa, Iowa City, IA, USA
[b] Civil and Environmental Engineering, University of Iowa, Iowa City, IA, USA
[c] Electrical and Computer Engineering, University of Iowa, Iowa City, IA, USA
* Corresponding author - msermet@uiowa.edu

**Abstract**
In recent years, advancements in artificial intelligence (AI) have led to the development of large language models like GPT-4, demonstrating potential applications in various fields, including education. This study investigates the feasibility and effectiveness of using ChatGPT, a GPT-4 based model, in achieving satisfactory performance on the Fundamentals of Engineering (FE) Environmental Exam. This study further shows a significant improvement in the model's accuracy when answering FE exam questions through noninvasive prompt modifications, substantiating the utility of prompt modification as a viable approach to enhance AI performance in educational contexts. Furthermore, the findings reflect remarkable improvements in mathematical capabilities across successive iterations of ChatGPT models, showcasing their potential in solving complex engineering problems. Our paper also explores future research directions, emphasizing the importance of addressing AI challenges in education, enhancing accessibility and inclusion for diverse student populations, and developing AI-resistant exam questions to maintain examination integrity. By evaluating the performance of ChatGPT in the context of the FE Environmental Exam, this study contributes valuable insights into the potential applications and limitations of large language models in educational settings. As AI continues to evolve, these findings offer a foundation for further research into the responsible and effective integration of AI models across various disciplines, ultimately optimizing the learning experience and improving student outcomes.

## 1. Introduction

The Fundamentals of Engineering (FE) exam is a nationally recognized exam in the United States that is designed to test the fundamental knowledge of engineering students and professionals (NCEES, 2023). The exam covers a wide range of engineering topics, including mathematics, chemistry, physics, statics, dynamics, and engineering economics (Barger et al., 2022). Passing the FE exam is a significant milestone for engineering students and is a requirement for becoming a professional engineer (PE) in the United States (Ressler, 2012). The FE exam serves as a benchmark for employers to evaluate the technical competence of engineering professionals and ensures that they possess the necessary skills and knowledge to practice engineering in a safe and effective manner (Koehn et al., 2008).

The Environmental Computer-Based Testing (CBT) Exam is a critical certification exam for environmental professionals in various fields, including engineering, geology, environmental science, and many others (Swenty and Swenty, 2022). Administered by the National Council of Examiners for Engineering and Surveying (NCEES), this exam evaluates an individual's competency in areas such as air quality, water quality, solid and hazardous waste management, environmental regulations, and sustainability (LaGro Jr, 2013). Passing the ENV CBT Exam is an essential requirement for obtaining a professional license and advancing one's career in the environmental industry (Bossart, 2020). The exam's rigorous nature and the breadth of knowledge required to pass make it an essential component of the professional development of environmental professionals.

As artificial intelligence (AI) continues to evolve and improve, it is playing an increasingly significant role in education and testing (Pedro et al., 2019). AI is being used to create intelligent tutoring systems, personalized study plans, and computer-based assessments that can adapt to the test-taker's individual learning needs (Sajja et al., 2023). Another area where AI can be utilized is in licensing examinations, such as the Fundamentals of Engineering (FE) Environmental CBT Exam, both in terms of assessing a test-taker's responses or taking the test itself. While the use of AI in licensing exams is promising, concerns have been raised about its potential impact on the credibility of these exams (Mbawke et al., 2023. Some argue that the use of AI within this context could lead to cheating and a lack of accountability (Arif et al., 2023). Others argue that AI can enhance the objectivity and fairness of the exams, particularly from the assessment standpoint, by eliminating human biases and subjectivity (Victor et al., 2023). In any case, with the increasing prevalence of prominent large language models (e.g., GPT-4) and AI chatbots (e.g., ChatGPT), the involvement of AI in such examinations needs thorough consideration and investigation.

This study explores the potential of ChatGPT to achieve satisfactory performance on the Fundamentals of Engineering (FE) Environmental Exam, with particular emphasis on the strength of different models for environmental engineering as well as the effectiveness of noninvasive prompt modifications in enhancing the model's accuracy. Additionally, the research highlights the substantial improvements in mathematical capabilities observed across successive iterations of ChatGPT models, demonstrating their aptitude for tackling complex engineering problems. By examining the performance of ChatGPT in the context of the FE Environmental Exam, this study sheds light on both the potential applications and challenges of large language models within educational settings. As the field of AI continues to progress, our findings serve as a basis for future research into the responsible and effective integration of AI models in various disciplines, ultimately aiming to enrich the learning experience and boost student outcomes. The paper further considers the concerns raised about the potential impact of AI on the credibility of licensing exams and highlights the need for further research and discussion on this topic.

### 1.1. Literature Review

The use of artificial intelligence (AI) in education has been rapidly increasing in recent years (Bai et al., 2022). One area where AI has shown potential is in helping students study for or pass exams, including standardized exams. This literature review aims to examine the current state of research on the use of AI tools in passing exams.

Several studies have investigated the use of AI tools in exam preparation. For example, Zhou et al. (2021) developed an intelligent online exam preparation system that used natural language processing and machine learning to analyze and provide personalized feedback on students' practice exams. Similarly, Huang et al. (2019) developed a mobile app that utilized AI algorithms to provide personalized recommendations on study materials and practice questions based on students' learning patterns. AI has also been used in exam assessment. Zhang et al. (2021) developed an AI-based exam grading system that used deep learning algorithms to grade multiple-choice questions with high accuracy. In another study, Yang et al. (2021) used AI to automatically generate exam questions that were comparable in difficulty to human-generated questions.

There have been several studies on the use of AI in passing exams, particularly in the field of medicine. For example, Zheng et al. (2021) developed an AI-based system that provided personalized exam preparation plans for medical students based on their learning progress and weaknesses. Similarly, Zhang et al. (2020) developed an AI-based system that helped medical students prepare for the United States Medical Licensing Examination (USMLE) by identifying knowledge gaps and providing targeted study materials. More recently, in the medical domain, several studies have investigated ChatGPT's performance on the USMLE. Kung et al. (2023) assessed ChatGPT's performance on all three USMLE exams, finding that the model performed at or near the passing threshold without any specialized training or reinforcement. Gilson et al. (2023) also evaluated ChatGPT's performance on USMLE Step 1 and Step 2 exams, concluding that the model achieved the equivalent of a passing score for a third-year medical student, making it a potential interactive medical education tool. Furthermore, Antaki et al. (2023) tested ChatGPT's accuracy in the ophthalmology question-answering space using multiple choice question banks for the Ophthalmic Knowledge Assessment Program (OKAP) exam. They reported encouraging results, but also suggested that domain-specific pre-training might be necessary to improve performance in ophthalmic subspecialties.

In the field of computer science, Bordt and Luxburg (2023) evaluated ChatGPT's performance in an undergraduate computer science exam on "Algorithms and Data Structures." They found that ChatGPT narrowly passed the exam, achieving 20.5 out of 40 points, indicating its potential in handling challenging tasks like university exams. However, the authors also noted that ChatGPT's training data might have included structurally similar questions, making it difficult to conclude whether the model truly understands computer science concepts.

In the context of business education, Terwiesch (2023) explored ChatGPT's performance in the final exam of a typical MBA core course, Operations Management. The study found that ChatGPT performed well in basic operations management and process analysis questions, including those based on case studies. However, the model made surprising mistakes in relatively simple calculations and struggled with more advanced process analysis questions. Despite these shortcomings, ChatGPT demonstrated a remarkable ability to modify its answers in response to human hints, suggesting its potential as a valuable tool in business school education.

These studies collectively highlight the promising capacity of ChatGPT in various educational domains, including computer science, medicine, ophthalmology, and business management. While its performance varies across disciplines and question types, continuous model development and domain-specific training could further enhance ChatGPT's effectiveness as an educational tool. As AI technology continues to advance, investigating the responsible and effective integration of large language models like ChatGPT across various disciplines becomes increasingly crucial (Mattas, 2023).

Despite the promising results of ChatGPT in various educational domains, there is a noticeable knowledge gap in the literature regarding its performance in the engineering field. Prior to the emergence of large language models, inference engines and AI systems (Sermet and Demir, 2018) that relied on domain ontologies (Sermet and Demir, 2019) were employed for intelligent assistance and question-answering in environmental sciences and hydrology fields (Yesilkoy et al., 2022). Engineering disciplines often involve the application of complex mathematical concepts and problem-solving skills, which may pose unique challenges for AI models like ChatGPT. Thus, it is crucial to investigate ChatGPT's capabilities in addressing engineering problems and its potential in assisting students and professionals in exam preparation and assessment.

Specifically, from an environmental engineering perspective, there is a lack of research on ChatGPT's performance in tackling questions related to this particular field. Environmental engineering is a multidisciplinary area that combines principles from various branches of engineering, such as civil, chemical, and mechanical engineering, as well as knowledge from environmental sciences, to address issues related to the environment and sustainability. Due to the breadth and complexity of environmental engineering topics, evaluating ChatGPT's proficiency in this domain can provide valuable insights into its applicability and limitations in addressing the diverse challenges posed by environmental engineering problems.

Furthermore, understanding ChatGPT's performance in the context of the Fundamentals of Engineering (FE) Environmental Exam, a critical certification exam for environmental professionals, can offer a more comprehensive perspective on its potential applications in the field. By bridging this knowledge gap, researchers and practitioners can better assess the suitability of AI models like ChatGPT in environmental engineering education and explore strategies for optimizing their performance and integration into educational settings.

In summary, there is a need to address the knowledge gap in the literature regarding ChatGPT's performance in engineering disciplines, particularly in environmental engineering. Investigating the model's capabilities in this domain will not only help in understanding its potential applications but also contribute to the responsible and effective integration of large language models in various educational contexts.

## 2. Methodology
### 2.1. Scope and Objectives
The purpose of this study is to provide insights into the potential benefits and limitations of utilizing AI in exam preparation and assessment. The stakeholders include engineering students, professionals, and licensing organizations such as the National Council of Examiners for Engineering and Surveying (NCEES). The value of this study lies in its potential to improve the efficiency and accessibility of exam preparation and assessment, as well as providing a foundation for further research and discussion on the use of AI in licensing exams. The research questions this study aims to address include: (1) How accurate is ChatGPT in answering FE Environmental CBT exam questions? (2) How feasible is it to utilize ChatGPT for exam preparation and assessment? (3) What are the potential benefits and concerns of using AI in licensing exams such as the FE Environmental CBT Exam?

### 2.2. AI Background
Artificial Intelligence (AI) has been making substantial strides in the field of engineering, with its applications in data analysis (Krajewski et al., 2021), data augmentation (Demiray et al., 2021), synthetic data generation (Gautam et al., 2022), prediction (Sit et al., 2021), and image analysis (Li and Demir, 2023) revolutionizing the way we approach complex problems. As AI continues to evolve and expand its reach, new horizons are being explored beyond the traditional realms of engineering. One such groundbreaking development is the emergence of large language models and conversational AI models like ChatGPT, which are reshaping the landscape of engineering education and communication. These AI-driven conversational agents not only have the potential to enhance the way we interact and engage with information, but also promise to democratize knowledge and revolutionize the learning experience for individuals across the globe.

ChatGPT, or GPT (Generative Pre-trained Transformer) for Chat, is a conversational AI model that has been trained on large amounts of text data using unsupervised learning techniques (OpenAI, 2023). It is an extension of the GPT-3 model and is capable of generating natural language responses to a variety of prompts. ChatGPT has demonstrated impressive results in various natural language processing tasks, including language translation, question-answering, and summarization. Compared to previous AI solutions, ChatGPT's advanced transformer architecture enables it to process longer and more complex sequences of text, leading to more accurate and coherent responses. However, like any AI model, ChatGPT is not perfect and may still make mistakes or produce biased responses, and its accuracy may vary depending on the specific task or data it is applied to.

Building on the success of GPT-3, OpenAI introduced GPT-3.5 and GPT-4 as subsequent iterations of ChatGPT, aiming to address some of the limitations and improve the model's overall capabilities. GPT-3.5, also known as GPT-3.5-turbo, has shown enhanced performance and efficiency compared to its predecessor, GPT-3.5-legacy, particularly in solving complex mathematical problems and providing more accurate responses to engineering-related questions (OpenAI, 2023). GPT-4, the most recent iteration of ChatGPT, brings further improvements in various aspects of natural language processing, including better context understanding, enhanced mathematical capabilities, and reduced bias in the generated responses. This version also introduces a "no vision" variant, which has been specifically optimized for text-based tasks without relying on visual information (OpenAI, 2023b).

These advancements in ChatGPT's capabilities have opened up new avenues for its application in diverse fields, including education, where it can be used to support learning, assessment, and knowledge acquisition. However, it is crucial to recognize that each version of ChatGPT comes with its own set of limitations and challenges. As a result, researchers and practitioners need to carefully consider the appropriateness of each model for their specific tasks and contexts, as well as continuously monitor and evaluate their performance to ensure the responsible and effective use of AI models in educational settings as well as with standardized tests.

### 2.3. Exam Overview and Data Collection

The Fundamentals of Engineering (FE) Environmental Exam, administered by NCEES, a nonprofit organization comprised of U.S. engineering and surveying licensing boards, is a comprehensive test that evaluates an individual's understanding of various environmental topics. The exam consists of 110 questions, which are designed to test the candidate's knowledge of a broad range of environmental engineering topics, including air quality, water quality, solid and hazardous waste management, environmental regulations, and sustainability (NCEES, 2020). In addition to traditional multiple-choice questions with one correct answer, the exam utilizes common alternative question types, including multiple correct options, point and click, drag and drop, and fill in the blank. The exam may also include questions that require the use of a calculator or specialized reference materials. Additionally, the exam may include non-scored questions, which are used for research purposes and do not affect the candidate's final score.

The dataset used in this study represents an unpublished practice exam that was collected from a variety of sources and was prepared by the University of Iowa Civil and Environmental Engineering department faculty. The majority of the questions came directly from FE Exam Prep books, with a number of materials available for faculty to use as references. The NCEES offers a PDF version of their exam prep book online, while other copyrighted materials are not available online and require separate subscriptions. While the FE Environmental practice exam preparation was originally started around 2015 when the NCEES switched to Computer-Based Testing (CBT), it was completed in Summer 2021. The final curation of the dataset utilized the aforementioned materials followed by individual faculty members being assigned to each section, with a leading faculty member coordinating the effort and contributions.

The audience for this dataset is primarily undergraduate students who are preparing to take the FE Environmental Exam, usually one semester prior to their graduation. However, it should be noted that the number of problems in each area is not representative of an actual exam, and the dataset is meant to serve as a resource for students to find problems, see worked-out solutions, and follow a video for each problem-solving method. In total, the preparation exam in the dataset consists of 134 questions. Out of these 134 questions, 1 was a drag and drop, 11 were fill in the blank, and the remaining ones were multiple-choice questions. Table A1 provides the number of questions and the weight of each section in the actual exam in comparison to the question distribution in our dataset. Figure 1 provides a comparative look at the question distribution in the dataset compared to the range of distribution in the FE Environmental Exam.

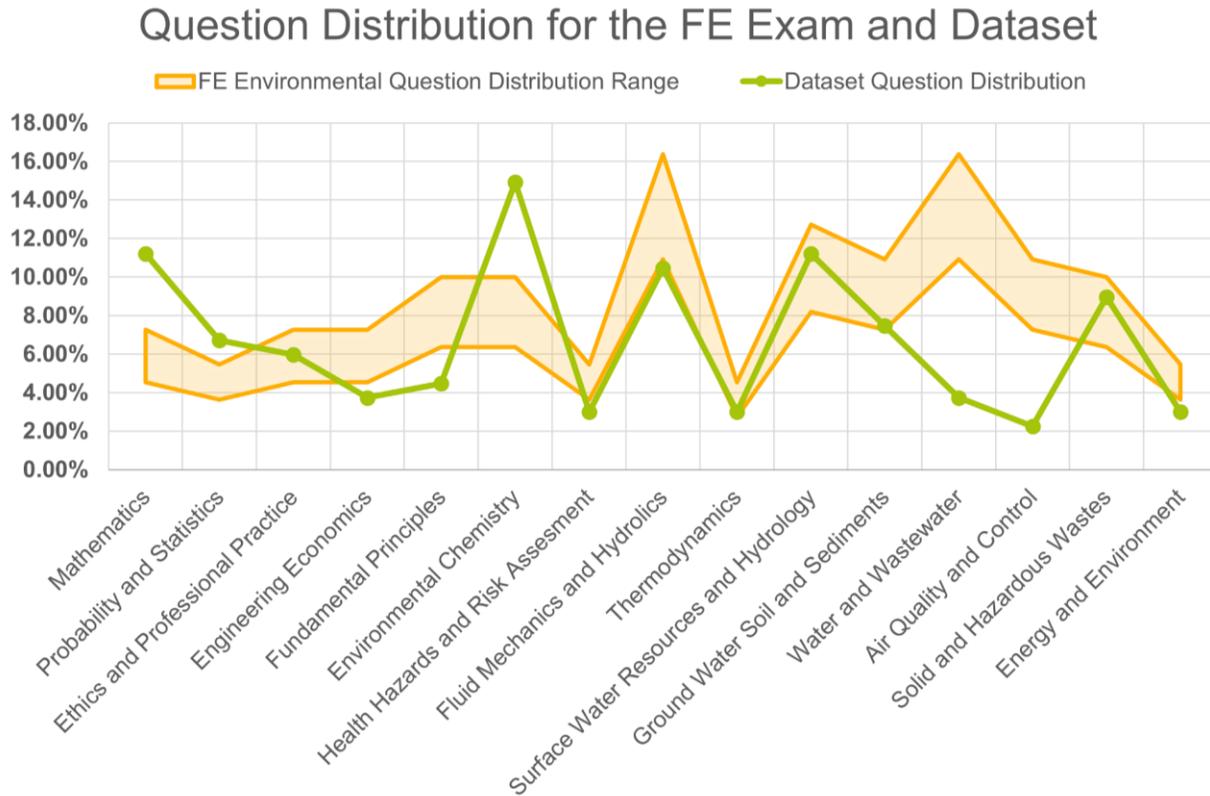

Figure 1: A comparison of question distribution in the dataset and the FE Environmental Exam, illustrating the degree of alignment between the two sets of questions and highlighting the study's representation of the exam's subject matter coverage.

### 2.4. Study Design

This section outlines the methodology employed to assess the performance of ChatGPT in answering questions from the Fundamentals of Engineering (FE) Environmental Exam. Two test sets were prepared based on the dataset, with each set containing a different approach to question presentation.

The first test set comprised verbatim questions, with no modifications except for addressing formatting issues that arose when transferring from the source PDF file. These questions were directly provided as prompts to ChatGPT, and were tested against GPT-3.5-Legacy, GPT-3.5-Turbo, and GPT-4 models. The second test set involved noninvasively refined questions, wherein additive guidance was incorporated without altering the question's content. For example, phrases were added to restrict the model to choosing a response only from a provided list, rather than generating a new one. This test set is tested against only the GPT-4 model to showcase the impact of noninvasive prompt modification. Finally, a separate experiment was conducted on GPT-3.5-Turbo, using the first test set to measure the effect of using different temperature values, which influences the randomness of the model's output. To avoid complications and potential interference in the analysis, each question was asked in unique calls. In summation of the methodology, Figure 2 describes the research workflow for the proposed assessment study.

To ensure a reproducible and efficient experiment process, a Python application was developed using the OpenAI API. The application implements functionality to process the test sets and conduct the defined experiments, saving all results for recordkeeping. A separate component, the autograder, was developed to automatically determine the correctness of the provided responses by cross-checking the questions with their respective answer keys. The autograder component utilizes GPT-3.5-Turbo as its underlying model for evaluation. Autograded results were then manually validated to ensure accurate labeling of correct and incorrect answers. Finally, all results were qualitatively investigated to derive meaningful insights.

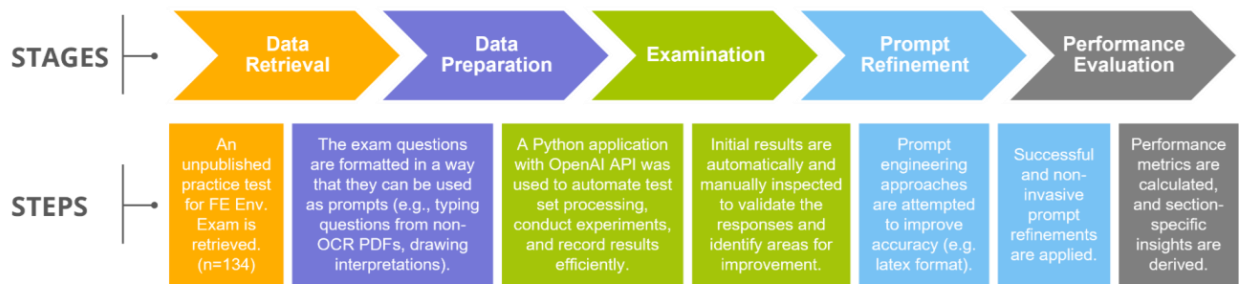

Figure 2: Research workflow for ChatGPT's performance assessment for FE Environmental Exam.

## 2.5. Study Limitations

The Fundamentals of Engineering (FE) exam is scored based on the number of correct answers selected by the examinee, with no deductions for wrong answers. The score is then converted to a scaled score, which takes into account any minor differences in difficulty across different exam forms. This scaled score represents the examinee's ability level and is compared to the minimum ability level for that particular exam, which has been determined by subject-matter experts using psychometric statistical methods. However, NCEES does not publish a passing score, and there is no predetermined percentage of examinees who should pass or fail the exam. All exams are scored in the same way, with first-time takers and repeat takers graded to the same standard. Therefore, there is no tangible pass/fail score for the FE exam. Hence, this study cannot claim for a fact that ChatGPT is or is not able to pass the FE Exam.

Although NCEES provides a diagnostic report to test-takers to convey which sections they performed well and which sections they require a better understanding, these reports are not public and a study that provides an analysis for such reports for the Environmental Engineering area could not be found in the literature. Thus, a section-specific comparison of success between human test takers and ChatGPT could not be made.

There are several other limitations to this study. Firstly, the dataset used for testing ChatGPT's performance may not be considered large or completely representative in terms of the question distribution. Also, it is unclear how representative the questions were of actual FE Exam questions despite the fact that they were curated by domain experts for educational and practice purposes. For example, the dataset lacks point and click question types, which, given ChatGPT's current absence of visual input capabilities, would render answering such questions infeasible. Additionally, there is a possibility of some, if any, questions in the dataset can be found online. Even though the dataset itself has never been released to the public, it is a possibility that students post some questions to online platforms such as Chegg for assistance. Thus, the probability of ChatGPT being aware of some questions in the dataset is nonzero.

Furthermore, ChatGPT is not a definitive source of factual information, as its primary purpose is to generate text that is contextually appropriate in response to user inputs. Secondly, the model's output is extremely sensitive to variations in the prompt's grammar, punctuation, and phraseology. Consequently, minor modifications to the input may have a significant effect on the generated response, necessitating cautious formulation. Third, the model's responses are intrinsically random, resulting in diverse responses to similar or identical prompts. Due to the lack of deterministic responses, the model's output should be evaluated with caution and not considered definitive. Hence, although the numbers reported in this study may potentially fluctuate upon reproducing the experiment, the calculated accuracy rates can reasonably be expected to be representative and insightful.

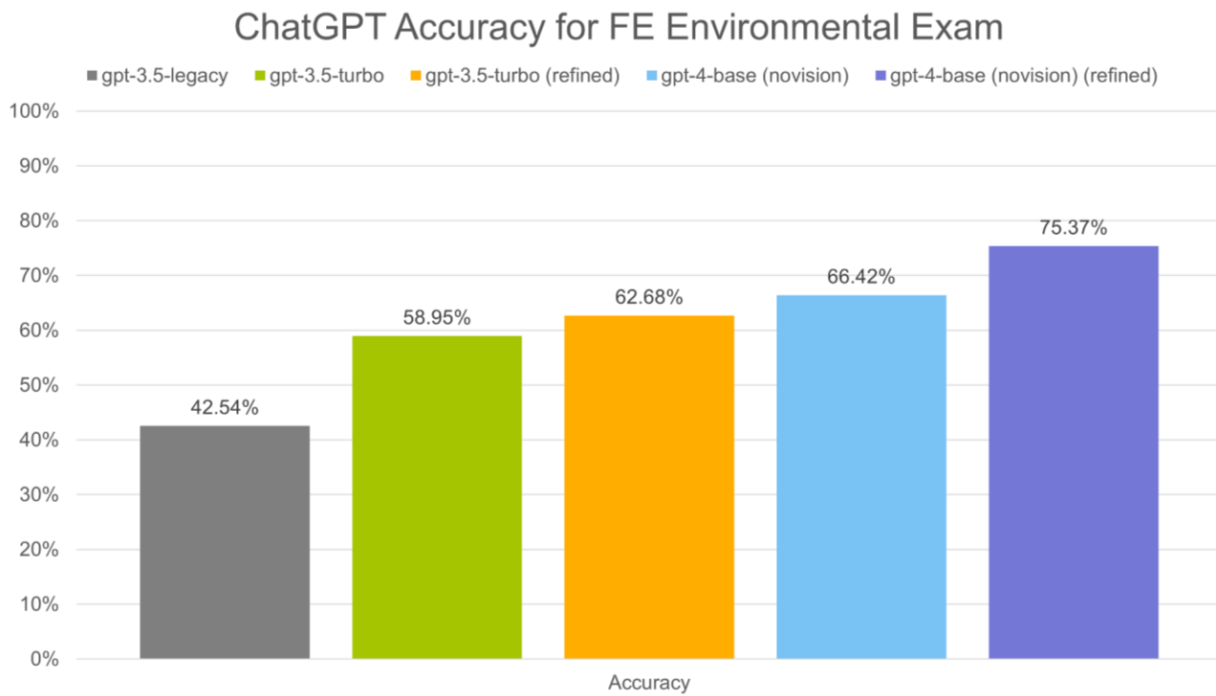

Figure 3: Overall accuracy of ChatGPT for the FE Environmental Exam. (n=134)

### 3. Results

In accordance with the objectives of the study, this section presents the results of the experiment that aimed to evaluate the accuracy and feasibility of using ChatGPT to pass the Fundamentals of Engineering

(FE) licensing exam. The results are presented through different graphic representations of the data that have been considered necessary to support our findings. In addition, the results are presented specific to the FE Exam sections to provide a more detailed analysis of ChatGPT's performance.

The results of our study indicate that ChatGPT exhibited varying levels of performance across different sections of the exam. Although a quantitative comparison to human test-takers who had a background in environmental engineering may not be made, the results showed that ChatGPT (GPT-4 Base Model - No Vision) achieved an overall accuracy of 66.42% on the dataset herein (Figure 3). When applying a refined approach to the same GPT-4 model, the accuracy improved to 75.37%, which can be reasonably considered a passing grade (Asghar, 2023). It is important to note that the questions were not invasively modified specifically for ChatGPT, and the model was not trained on the dataset beforehand.

The section-specific accuracy of ChatGPT (Figure 4) provide valuable insights into the model's strengths and weaknesses across various topics in the FE Environmental Exam. From the results, it is evident that ChatGPT performs exceptionally well in certain sections, such as Ethics and Professional Practice, Thermodynamics, and Probability and Statistics, particularly when using the GPT-4 No Vision model with refined prompts. This suggests that ChatGPT can be a reliable tool for tackling questions in these areas. On the other hand, the model's performance in sections like Ground Water Soil and Sediments and Surface Water Resources and Hydrology remains relatively low, even with the refined prompts. This indicates potential limitations in ChatGPT's understanding of these specific subject matters and highlights the need for further improvement in the model or targeted training to enhance its performance in these areas.

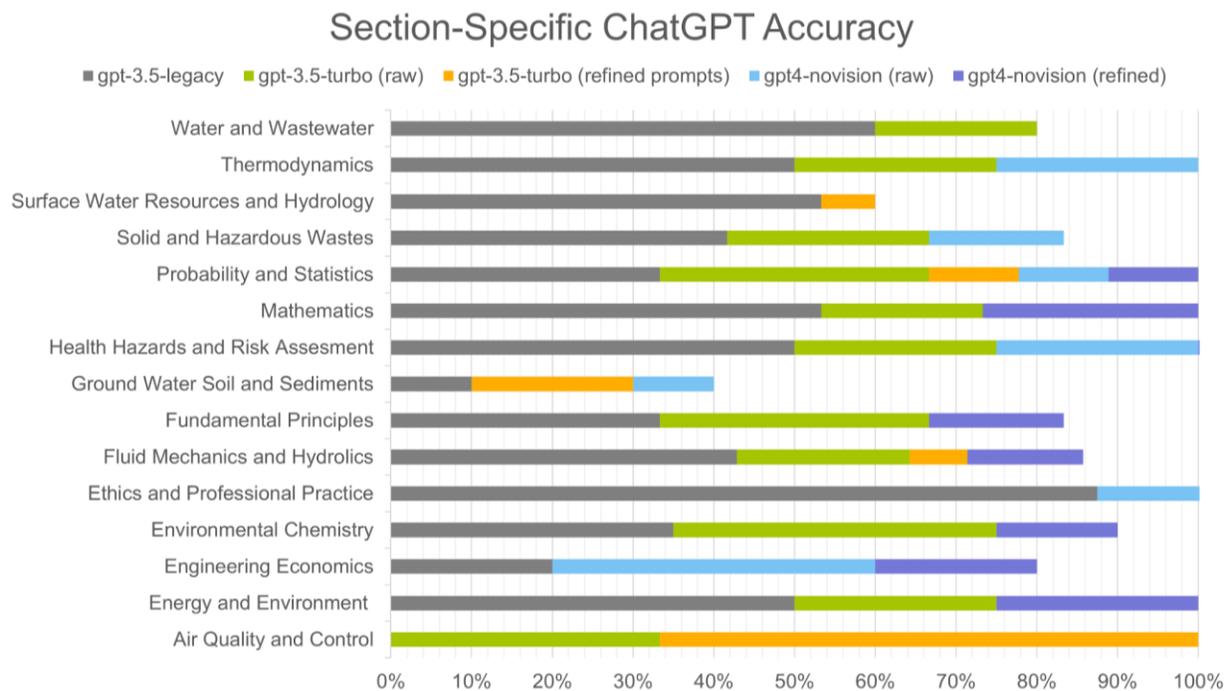

Figure 4: Section-specific ChatGPT accuracy for different models and prompt refinements. (n=134)

### 3.1. Effects of Temperature Configuration in Engineering Questions

Understanding the impact of temperature configuration on ChatGPT's performance is crucial in the context of the FE Environmental Exam. Here, we present the results of our analysis using the GPT-3.5-Turbo model with varying temperature values. The temperature parameter, which ranges from 0 to 1, influences the randomness of the model's output. Higher temperatures (closer to 1) yield more diverse responses, while lower temperatures (closer to 0) produce more focused and deterministic outputs.

Figure 5 shows the section-specific accuracy of the GPT-3.5-Turbo model for different temperature values across all 134 questions in the dataset. From the results, it is evident that there is a variation in accuracy based on the temperature parameter. The highest accuracy (i.e., 0.59) was achieved with a temperature setting of 0.75, while the lowest accuracy (i.e., 0.47) was observed at a temperature setting of 0.5. The accuracies for the other temperature settings were found to be relatively close, with *TEMP*=1 at 0.56, *TEMP*=0.25 at 0.53, and *TEMP*=0 at 0.56.

These findings suggest that the choice of temperature setting plays a significant role in the performance of ChatGPT when answering FE Environmental Exam questions. To optimize ChatGPT's performance for various sections or topics in the exam, it is essential to identify and use the appropriate temperature setting based on the specific context. Overall, our analysis indicates that a temperature setting of 0.75 is the most suitable for achieving the best accuracy in this particular application. However, further investigation and experimentation with different temperature values might be necessary for other contexts or specific sections of the exam.

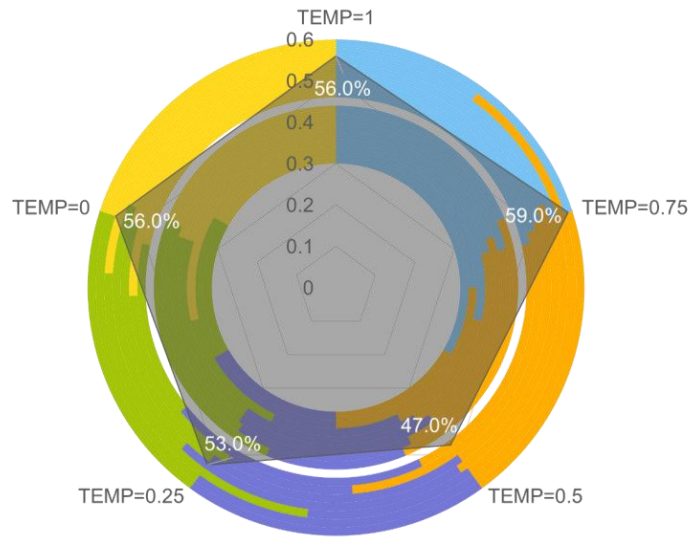

Figure 5: Section-specific GPT-3.5-Turbo accuracy for different temperature values. (n=134)

### 3.2. Utility of Noninvasive Prompt Refinements

In this study, several noninvasive refinement approaches were explored to determine which methods consistently improved the performance of ChatGPT in answering FE Environmental Exam questions. The following approaches were investigated:

*Approach 1.* Utilizing LaTeX formatting to present questions or requiring ChatGPT to use LaTeX in its responses did not yield any significant improvements in the model's performance.

*Approach 2.* Transcribing visual elements, such as diagrams, into textual descriptions for ChatGPT. Only two questions in the dataset (one from Mathematics and one from Ground Water Soil and Sediments) contained unique information in their diagrams that was not present in the question text itself. For both scenarios, describing the diagrams textually did not lead to consistently accurate results.

*Approach 3.* Providing a system message instructing ChatGPT to focus on providing clear and concise explanations for its answers, and to make an educated guess from the provided options when in doubt, similar to how an actual student would approach the questions. This third approach proved beneficial.

Consequently, the only question refinement that was applied was Approach 3 by providing a system message as follows: "*Your job is to provide explanations and answers to the best of your ability, with clear and concise explanations. If the question is a multiple-choice question, provide the most suitable option with an explanation. If the question does not have multiple-choice options, provide the calculated answer with a clear explanation. Here is the question:*" This message encourages the model to justify its solution and make an educated guess from the provided options when appropriate.

One advantage of this refinement approach is that it prompts the model to return an explanation for its answers, enabling the verification of the correctness of its reasoning. Without the refinement, ChatGPT may simply return the correct option without explanation. Interestingly, for one question in the Ethics section, GPT-4 consistently provided the correct answer when asked verbatim but never offered an explanation. However, with the refinement, the model always explained its reasoning but arrived at the wrong answer. This phenomenon suggests that GPT-4 may have had prior knowledge about that particular question and knew the correct answer, but when asked to solve it independently, it lacked the necessary underlying knowledge to correctly answer the question. This highlights the importance of refining prompts to ensure more reliable and insightful results from ChatGPT.

Upon implementing noninvasive prompt modifications to the GPT-4 model (no vision), our results indicate a significant improvement in accuracy when answering questions from the FE Environmental Exam dataset (Figure 6). We observed that refinements had a more substantial impact on higher temperatures, which is why we conducted the experiment at a temperature setting of 0.75.

The experimental group, utilizing the modified prompts, demonstrated a mean accuracy percentage point increase of 8.95% compared to the control group that employed unmodified prompts. This measurable utility was consistently observed across various question types, including those addressing theoretical concepts, problem-solving tasks, and practical applications. Furthermore, a detailed analysis of the response patterns revealed that the noninvasive prompt modifications effectively guided the model toward more concise and accurate answers, mitigating the likelihood of generating irrelevant or ambiguous responses. These findings substantiate the efficacy of prompt modification as a viable approach to enhance the performance of the GPT-4 model in educational contexts, specifically for the preparation of the FE Environmental Exam.

It is also noteworthy that for some sections, such as Engineering Economics, Environmental Chemistry, and Mathematics, the application of refined prompts led to a significant improvement in the model's accuracy. Refinements showed no particular improvement in sections with textual content; however, they provided somewhat of an improvement in math-heavy sections. We observed that, with the refinement, the model does not produce a new response that is not present in the options, nor does it simply choose the closest numerical values even if they are highly irrelevant. Instead, it tends to work towards one of the provided options and makes an educated guess. This behavior further emphasizes the value of noninvasive prompt refinements in guiding ChatGPT towards more accurate, reliable, and justifiable responses, particularly in the context of the FE Environmental Exam.

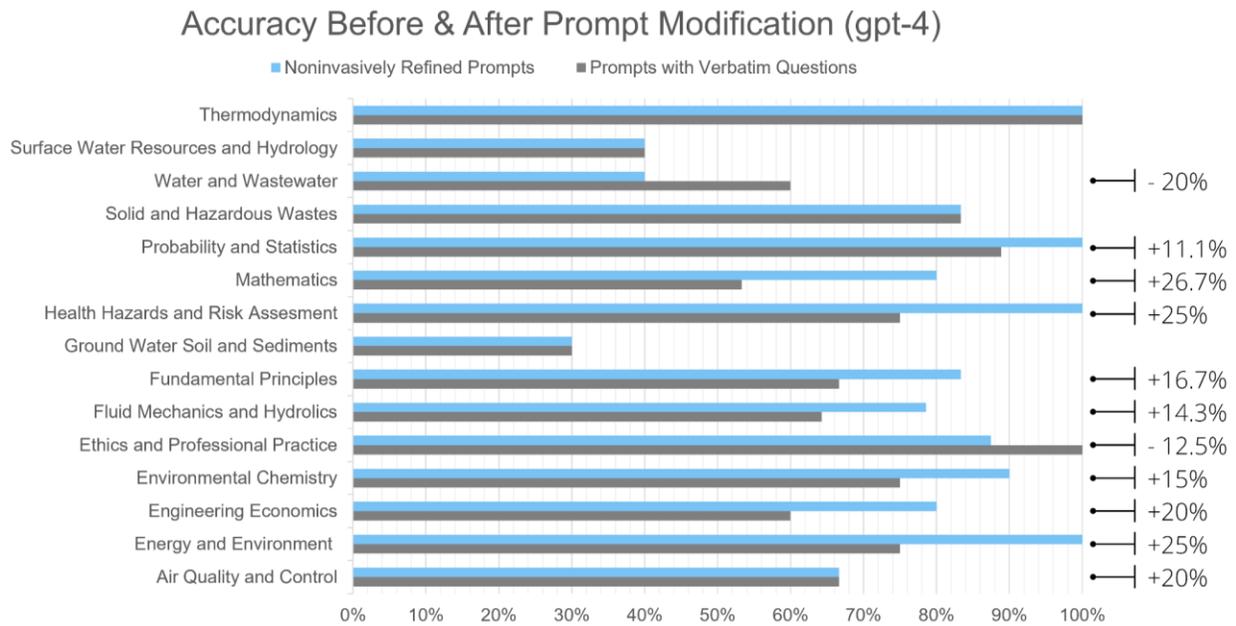

Figure 6: A comparative bar chart to reflect the change in accuracy for each section with and without prompt refinement. (n=134)

Based on the results and the impact of refined prompts, several patterns can be observed. Firstly, the refined test set led to a noticeable improvement in numerous sections, such as Energy and Environment (25%), Engineering Economics (20%), Environmental Chemistry (15%), Fluid Mechanics and Hydraulics (14.29%), Fundamental Principles (16.67%), Health Hazards and Risk Assessment (25%), Mathematics (26.67%), and Probability and Statistics (11.12%). This demonstrates the effectiveness of noninvasive prompt refinements in enhancing ChatGPT's performance across various topics.

Secondly, for certain sections like Air Quality and Control, Ground Water Soil and Sediments, Solid and Hazardous Wastes, Surface Water Resources and Hydrology, and Thermodynamics, the application of refined prompts did not yield any improvement. The performance in these sections may be limited due to the need for detailed and vivid descriptions of images in questions, as well as the complexity of questions that involve multiple topics. Additionally, the model might struggle with questions involving complex hydrogeological settings, multiple processes, and variable parameters that require detailed and specific information beyond the question text for providing accurate and helpful responses.

Interestingly, the refined test set resulted in a decrease in performance for Ethics and Professional Practice and Water and Wastewater sections. This could likely be attributed to the fact that the refinement minimizes lucky guesses and requires educated and justifiable attempts at solving the question, as described earlier in this section.

Therefore, it is essential to recognize that the optimal refinement strategy may not be universally applicable across all sections, and further investigation is required to develop tailored refinements for each section or topic. Overall, however, it can be claimed that the refinement increases the actual correctness of the responses provided by the model.

### 3.3. Reliability and Mathematical Capability of ChatGPT for Engineering Questions

Our comparative analysis of the mathematical capabilities of different generations of ChatGPT models reveals significant improvements in their ability to tackle engineering questions. Figure 7 displays the weighted accuracy for math-heavy sections of the FE Environmental Exam, including Math, Probability and Statistics, and Engineering Economics. Our findings show that GPT-3.5-turbo exhibits superior mathematical capabilities compared to its predecessor, GPT-3.5-legacy, with an observed accuracy increase from 41.38% to 62.07%. When refining GPT-3.5-turbo, the accuracy further improves to 65.51%. The most recent iteration, GPT-4-base (no vision), demonstrates comparable performance to the refined GPT-3.5-turbo with an accuracy of 65.51%. However, when refining GPT-4-base (no vision), the model significantly outperforms its predecessors, achieving an impressive accuracy of 86.21%. These advancements across successive model versions suggest that the developers have addressed limitations in the AI's mathematical problem-solving abilities to a considerable extent. The improvements in the mathematical capabilities of ChatGPT models offer promising potential for their application in educational and professional engineering contexts, where accurate and efficient problem-solving is essential.

Despite the improvements observed in ChatGPT's mathematical capabilities, our analysis of its performance on the FE Environmental Exam has unveiled considerable disparities in accurately answering questions across various exam sections. While the refined GPT-4-base (no vision) model showcased a robust understanding of theoretical concepts and the ability to construct suitable formulas for problem-solving, its overall mathematical capability proved inadequate for complex engineering calculations. This limitation resulted in fluctuating accuracy, particularly in sections like Engineering Economics and Mathematics, where a series of calculations are necessary to arrive at correct answers.

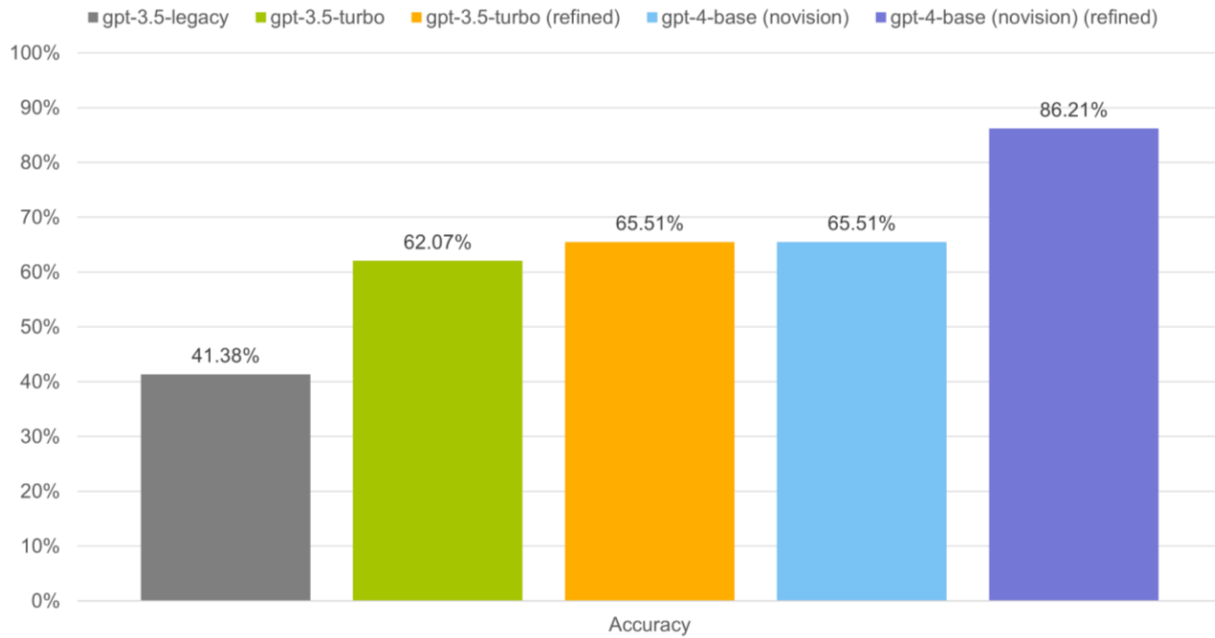

Figure 7: Weighted Accuracy of ChatGPT for the FE Environmental Exam Math-Heavy Sections (n=29)

Our findings indicate that the refined GPT-4-base (no vision) model excelled at understanding and constructing appropriate formulas. However, its mathematical capabilities were insufficient for accurately solving problems, especially in the Engineering Economics and Mathematics sections. This inconsistent performance highlights ChatGPT's mathematical capability as a significant constraint when applied to calculation-intensive engineering problems. In several instances, the refined GPT-4-base (no vision) model successfully identified the relevant formulas and concepts but failed to accurately perform the required calculations, leading to incorrect answers despite its correct understanding of the underlying principles. These errors can be traced back to the model's limitations in handling multi-step calculations and intricate mathematical operations, which are often vital in solving engineering problems.

In addition to these inconsistencies, our analysis also unveiled notable variability in ChatGPT's performance concerning the comprehension of mathematical relationships. Interestingly, GPT-3.5 occasionally provided correct answers, even when its calculations and overall formulation were incorrect. This finding suggests that, in some cases, GPT-3.5 did not genuinely solve these questions but rather stumbled upon the correct answers in multiple-choice questions. Consequently, although the overall accuracy for GPT-3.5 seems reasonable, such fortuitous outcomes emphasize the model's limitations in consistently and reliably solving complex mathematical problems.

4.  Discussion and Conclusions

In conclusion, our study demonstrates the potential of ChatGPT as a valuable tool for addressing engineering problems and preparing for the Fundamentals of Engineering (FE) licensing exam. The GPT-4 base model with no vision achieved a notable overall accuracy of 66.42% on the FE Environmental Exam dataset, indicating its capability to provide satisfactory results across various question types (e.g., multiple-choice, fill in the blank). Our research also highlights the significant impact of noninvasive prompt modifications on the GPT-4 base model (no vision), leading to a mean accuracy increase of 8.95% for the FE Environmental Exam, resulting in an overall accuracy of 75.37%. While limitations were observed in specific subject areas, such as Ground Water Soil and Sediments, these findings emphasize the importance of continuous model development and prompt refinement to effectively address complex questions.

Moreover, the study's comparative analysis of different generations of ChatGPT models underscores the significant improvements in mathematical capabilities for engineering applications. The GPT-4 (no vision) model demonstrated superior performance compared to its predecessors, illustrating the developers' successful efforts in addressing the AI's mathematical limitations to some degree. These advancements in ChatGPT models hold promising implications for their use in educational and professional engineering settings, where accurate problem-solving is of paramount importance.

Furthermore, it is essential to consider that with dedicated training on large language models such as GPT-4, AI models can realistically be expected to achieve expert proficiency in the future on licensing exams, regardless of their difficulty. This suggests that as AI technology continues to advance, the potential of large language models to provide reliable and accurate assistance in exam preparation will only increase, ultimately revolutionizing the way students and professionals approach these high-stakes assessments. Finally, it should be noted that GPT-4 has the capability to interpret visual inputs, although not publicly available at the time of this publication, which will particularly be of utility for the engineering domain where schematics and drawings can play a major role.

Given that the FE Environmental exam is highly competitive and only 64% of participants passed it in 2023 (NCEES, 2020), the fact that an unrefined, domain-independent, widely-available AI can potentially get a passing grade in an exam that is supposed to assess and prove the credibility and merits of future engineers is nontrivial and require deep further discussion. The use of AI models in passing the FE Environmental licensing exam has the potential to both positively and negatively impact the exam and its participants. On the positive side, AI models such as ChatGPT can assist individuals in studying for the exam, potentially increasing their chances of passing. AI models can provide personalized study plans and tutoring systems that can adapt to the test-taker's individual learning needs, ultimately resulting in better exam preparation. Furthermore, AI models can improve the efficiency and accessibility of the exam by reducing the time and resources required for traditional exam preparation methods. Additionally, AI can provide objective and unbiased assessment of exam takers' responses, reducing the risk of human bias and subjective grading.

However, the use of AI models in the exam also raises concerns about their potential impact on the credibility of the exam. AI models can pass the exam without any training or customization, leading to concerns about cheating and a lack of accountability. Furthermore, there may be concerns about the validity and reliability of AI-generated responses, as well as the possibility of errors or inaccuracies in the training data, which is particularly problematic from an education and assessment standpoint. Therefore, there is a need for careful consideration and investigation of the use of AI models in licensing examinations such as the FE Environmental exam to ensure that they do not undermine the credibility of the exam or introduce new challenges to student preparation.

Furthermore, the observed limitations in ChatGPT's mathematical capability raise concerns about the reliability of using the model for problem-solving purposes in engineering education. While the model's understanding of theoretical concepts is promising, its inability to consistently deliver accurate calculations may impede its practical application in educational settings. Students and educators seeking assistance from AI models like ChatGPT for solving complex engineering problems may face challenges due to these limitations, potentially leading to misconceptions and decreased learning outcomes.

### 4.1. Future Work and Opportunities

In light of the promising results and potential applications of AI models like GPT-4 in educational contexts, several avenues for future research and development can be considered:

a) **Integrating computational tools and engineering software with AI:** One promising direction is the development and integration of plugins, which has already begun to be explored by OpenAI with partners and early adopters at the time of writing. To address potential inconsistencies in expert knowledge areas and fields that require access to specialized tools, ChatGPT can be integrated with external tools. For instance, although ChatGPT was able to provide correct representations of formulas, it occasionally made mistakes in the actual calculations. However, when prompted to write Python code to execute the calculations, it successfully provided the correct answer to the exam question. This highlights the potential benefits of integrating tools such as code interpreters and platforms like Wolfram Alpha. Making these integrations available to the public could potentially propel ChatGPT to substantially higher exam performances and increase its viability for use in engineering education. Future research should explore the development of efficient and user-friendly plugin systems to facilitate seamless interaction between AI models like ChatGPT and specialized tools, ultimately enhancing the educational experience and outcomes for learners across various disciplines.
b) **Developing and integrating AI study partners into the classroom:** ChatGPT allows a use case for students to prepare for the FE exam, even in GPT-4's current form without training or prompt refinement. One of the benefits of using ChatGPT is that it provides detailed explanations for practice questions, which can help students improve their understanding of the material and assist with problem-solving. Additionally, ChatGPT can be used to generate variations of practice questions based on provided examples, allowing students to practice applying their knowledge to different scenarios. By using ChatGPT in conjunction with other study materials, students can gain a deeper understanding of the concepts covered on the FE Environmental Exam and improve their chances of success on the exam.
c) **Addressing AI challenges in education:** There is a need for continued research and development in the use of AI in exam preparation and assessment, with a focus on addressing the challenges and potential limitations of such approaches. One potential future research direction is to develop AI models that can effectively detect and mitigate cheating in exams, particularly in computerized tests. This can be achieved by incorporating features such as keystroke analysis, biometrics, and facial recognition to detect irregularities during the exam. Additionally, there is a need for developing and refining AI models that can assist in exam preparation by providing personalized study plans, generating practice questions, and providing feedback on performance. This can potentially improve the effectiveness of exam preparation and increase the likelihood of passing the exam.

d) **Investigating the role of AI in learning**: Another area that requires further research is the potential impact of AI on the credibility and fairness of licensing exams. As AI tools become more sophisticated, there is a risk of increased cheating and a lack of accountability in the exam process. Therefore, future studies can investigate ways to mitigate these risks while still leveraging the benefits of AI in exam preparation and assessment. Additionally, case studies can be conducted to evaluate the effectiveness of AI models in improving student performance on licensing exams and their potential for use in other educational contexts.

e) **Preventing cheating with AI-resistant exam questions**: With the increasing capabilities of AI models, it is essential to develop strategies to prevent cheating in high-stakes exams. Examination professionals can explore the design of custom questions that are specifically tailored to thwart AI models, which can informally be referred to as "Captcha for exam questions." By incorporating these AI-resistant questions into the assessment process, examination integrity can be maintained to some degree while still harnessing the benefits of AI in education.

f) **Preparing next-generation professionals for the age of AI**: It is crucial for students to be aware of the challenges associated with AI, such as potential biases in output, the necessity for continuous human oversight, and the risks of misuse of large language models. These challenges are not unique to education but are inherent to transformative digital technologies. Educators can play a vital role in ensuring that students are introduced to these issues early on, thus promoting responsible AI usage and fostering a deeper understanding of the societal implications of AI applications.

g) **Enhancing accessibility and inclusion**: AI models can contribute to a more inclusive learning environment by providing educational materials tailored to disadvantaged engineers or those for whom English is not their first language. By developing virtual tutors capable of communicating in various languages, the accessibility of study materials and support for the FE exam can be significantly improved, facilitating an equitable learning experience for diverse student populations.

**Acknowledgements**
The dataset used in this study was curated by the University of Iowa Civil and Environmental Engineering department faculty for an FE exam preparation course. ChatGPT was utilized in the preparation of this article; any generated text is reviewed and revised for accuracy.

**Data Availability**
Due to the copyrighted nature of the questions used in this dataset, the dataset provider was assured raw data would remain confidential and would not be shared. (Data not available / The data that has been used is confidential.)

**Author Contributions**
**Vinay Pursnani**: Software, Methodology, Data curation, Formal Analysis, Validation. **Yusuf Sermet**: Conceptualization, Writing - original draft, Methodology, Investigation, Validation, Visualization. **Ibrahim Demir**: Conceptualization, Writing – review & editing, Supervision, Project administration, Resources.

**Funding**
This research received no external funding.

**Appendix**

Table A1: Comparative view of section distributions in the dataset and the actual exam (NCEES, 2020).

| Exam Section | FE Environmental Exam (110) | | Our Dataset (134) | |
| --- | --- | --- | --- | --- |
| | Weight | # of Questions | Weight | # of Questions |
| Mathematics | 4.5-7.2% | 5-8 | 11.19% | 15 |
| Probability and Statistics | 3-4% | 4-6 | 6.72% | 9 |
| Ethics and Professional Practice | 5-7% | 5-8 | 5.97% | 8 |
| Engineering Economics | 4-5% | 5-8 | 3.73% | 5 |
| Fundamental Principles | 3-5% | 7-11 | 4.48% | 6 |
| Environmental Chemistry | 10-15% | 7-11 | 14.93% | 20 |
| Health Hazards and Risk Assessment | 5-7% | 4-6 | 2.99% | 4 |
| Fluid Mechanics and Hydraulics | 8-13% | 12-18 | 10.45% | 14 |
| Thermodynamics | 3-4% | 3-5 | 2.99% | 4 |
| Surface Water Resources and Hydrology | 9-14% | 9-14 | 11.19% | 15 |
| Groundwater, Soils, and Sediments | 8-13% | 8-12 | 7.46% | 10 |
| Water and Wastewater | 13-19% | 12-18 | 3.73% | 5 |
| Air Quality and Control | 9-14% | 8-12 | 2.24% | 3 |
| Solid and Hazardous Waste | 9-14% | 7-11 | 8.96% | 12 |
| Energy and Environment | 3-5% | 4-6 | 2.99% | 4 |